\documentclass[twocolumn,showpacs,showkeys,amsmath,amssymb]{revtex4}

\usepackage{graphicx}
\usepackage{dcolumn}
\usepackage{bm}
\usepackage[latin2]{inputenc}

\def\rmssm{$\slash \!\!\!\! R_p$--MSSM}
\begin{document}
\title {Renormalization group parameter evolution \\ of the minimal
  supersymmetric standard model with $R$-parity violation} 
\author{Marek Góźdź} 
\email{mgozdz@kft.umcs.lublin.pl} 
\author{Wiesław A. Kamiński} 
\email{kaminski@neuron.umcs.lublin.pl} 
\affiliation{Theoretical Physics Department, Maria Curie--Skłodowska
University, Lublin, Poland}
\begin{abstract}
A comparison of spectra obtained using the 1--loop MSSM and 2--loop
\rmssm \ renormalization group equations is presented. Influence of
higher loop corrections and $R$-parity violating terms is
discussed. Some numerical constraints on the $R$-parity violating
parameters are also given.
\end{abstract}

\maketitle
\section{Introduction}

The recent observations of neutrino oscillations \cite{SNO, Ahn, Det} and reported, although not confirmed, 
discovery of neutrinoless double beta decay \cite{0nu2beta} give strong motivation
for studying some aspects of physics beyond the Standard Model (SM). One
of the most popular approach to such a physics is the incorporation of supersymmetry,
which leads, in the simplest case, to the Minimal Supersymmetric
Standard Model (MSSM) \cite{ChoMisiak}. This framework is very
attractive for theorists since it is the simplest supersymmetric model
possible that leads to unification of the coupling constants. It also
provides a natural way to incorporate of lepton and baryon number violating
processes, which is impossible in SM.

It was widely approved to neglect the terms violating lepton and baryon
numbers in actual calculations, due to smallness of the relevant
coupling constants \cite{HirKlap, HaugVer, WodKam}. 
However, even if perfectly valid in low energies, this assumption could have lead to
errorneus predictions in high energy limits. We show by explicit
simulations that this approach is justified in most cases.

The paper is organized as follows. In the next section we present the
model. We also explain our conventions and notations here. In Section III the
results are presented and discussed. A short conclusion is given at the
end.

\section{Theory}

The minimal (in the particle and interaction content) supersymmetric
extension of SM means introducing a superpartner for each particle 
and adding a second Higgs doublet (together with its higgsinos).

We closely follow the conventions used in \cite{MartinVaughn,
ADedesD}. The MSSM superpotential can be written in the form
\begin{eqnarray}
W^{MSSM} &=& \epsilon_{ab} [(\mathbf{Y}_E)_{ij} L_i^a H_1^b \bar E_j 
+ (\mathbf{Y}_D)_{ij} Q_i^{ax} H_1^b \bar D_{jx} \nonumber \\
&+& (\mathbf{Y}_U)_{ij} Q_i^{ax} H_2^b \bar U_{jx} + \mu H_1^a H_2^b ], 
\label{wmssm}
\end{eqnarray}
where {\bf Y}'s are 3$\times$3 Yukawa matrices. $L$ and $Q$ are the
$SU(2)$ left-handed doublets while $\bar E$, $\bar U$ and $\bar D$
denote the right-handed lepton, up-quark and down-quark $SU(2)$ singlets,
respectively. $H_1$ and $H_2$ mean two Higgs doublets. We have
introduced color indices $x,y,z = 1,2,3$, generation indices
$i,j,k=1,2,3$ and the SU(2) spinor indices $a,b,c = 1,2$.

Since supersymmetry is not observed, we need to introduce some mechanism to
break it. The simplest possibility is the so-called soft breaking given by the
Lagrangian
\begin{eqnarray}
{\cal L}^{soft} &=& \epsilon_{ab} [(\mathbf{A}_E)_{ij} l_i^a h_1^b \bar e_j 
+ (\mathbf{A}_D)_{ij} q_i^{ax} h_1^b \bar d_{jx} \nonumber \\
&+& (\mathbf{A}_U)_{ij} q_i^{ax} h_2^b \bar u_{jx} + B \mu h_1^a h_2^b], 
\end{eqnarray}
where lowercase letters stand for scalar components of respective chiral
superfields, and 3$\times$3 matrices {\bf A} as well as $B\mu$ are
the soft breaking coupling constants. We introduce also a scalar mass term
of the form
\begin{eqnarray}
{\cal L}^{mass} &=& \mathbf{m}^2_{H_1} h_1^\dagger h_1 +  
                    \mathbf{m}^2_{H_2} h_2^\dagger h_2 + 
     q^\dagger \mathbf {m}^2_Q q + l^\dagger \mathbf {m}^2_L l \nonumber \\
&+&  u \mathbf {m}^2_U u^\dagger + d \mathbf {m}^2_D d^\dagger + 
     e \mathbf {m}^2_E e^\dagger.
\end{eqnarray}

Looking at (\ref{wmssm}) it is clearly visible, that $W^{MSSM}$ preserves a
discrete symmetry called $R$-parity and defined as 
$$
R_p = (-1)^{3B+L+2S},
$$
where $B$, $L$, and $S$ are the baryon, lepton and spin numbers,
respectively. From the definition follows that ordinary particles have
$R_p=+1$ whereas supersymmetric (SUSY) particles have $R_p=-1$. The
consequence of $R_p$ conservation is a stable lightest SUSY particle, which
can be a serious candidate for cold dark matter, and a stable proton.

However, nothing motivates theoretically $R_p$ conservation, especially
that in order to avoid rapid proton decay, at least the lepton number
parity $L_p = (-1)^{L+2S}$ {\it or} the baryon number parity $B_p =
(-1)^{3B+2S}$ must be conserved, {\it but not necessarily both}. It
follows that we can add new terms to the superpotential, which violate
separately the lepton and baryon number:
\begin{eqnarray}
W^{\not R_p} &=& \epsilon_{ab}\left[
\frac{1}{2}(\mathbf{\Lambda}_{E^k})_{ij} L_i^a L_j^b \bar E_k 
+(\mathbf{\Lambda}_{D^k})_{ij} L_i^a Q_j^{xb} \bar D_{kx} \right] \nonumber \\
&+& \frac{1}{2}\epsilon_{xyz}(\mathbf{\Lambda}_{U^i})_{jk}\bar U_i^x\bar
D_j^y \bar D_k^z + \epsilon_{ab}\kappa^i L_i^a H_2^b.
\end{eqnarray}
The full superpotential becomes now 
\begin{equation}
W = W^{MSSM} + W^{\not R_p}
\end{equation}
and defines a new model, often called \rmssm. It introduces nine new
Yukawa coupling constants but, as mentioned above, proton stability
requires to reject either lepton number or baryon number violating
terms. If one wants to apply this model to the description of the
neutrinoless double beta decay
$$
A(Z,N) \to A(Z+2,N-2) + 2e^-,
$$ 
in which the lepton number is violated by two units, the matrices
$\mathbf{\Lambda}_{U^i}$ have to be set equal to zero. In this paper we will
not do that in order to stay completely general and to be able to count
contributions coming from all sources.

All the parameters (coupling constants and masses) of \rmssm \ are running
parameters in the sense of renormalization group, that is they obey a
set of equations of the form
\begin{equation}
\mu \frac{dx}{dt} = \frac{1}{16\pi^2}\beta^{(1)}_x + 
\frac{1}{(16\pi^2)^2}\beta^{(2)}_x + \dots,
\end{equation}
where $t=\log E$ and $E$ is the renormalization scale; $\beta^{(n)}_x$
is the $n$--loop renormalization group beta function for the parameter
$x$. Our previous calculations \cite{WodKam, WodPagerka, WodSimkovic}
were based on 1--loop MSSM renormalization group equations because we
roughly estimated that the corrections coming from $R$-parity violation are of
the order of few percent. We check this assumption in detail in the
next section.

\section{Results}

\begin{table}

\caption{\label{tab1} Mass matrices of supersymmetric particles (third
family) and values of the coupling constants, calculated using 1--loop
MSSM and 2--loop \rmssm\ renormalization group equations. All
supersymmetric masses are unified (500 GeV) at the GUT scale $10^{16}$
GeV. Other MSSM parameters were $\tan\beta = 5$ and positive sign of
$\mu$.}
\begin{ruledtabular}
\begin{tabular}{ccccc}
 & 1--loop GUT & 2--loop GUT & 1--loop $m_Z$ & 2--loop $m_Z$\\
 & MSSM & \rmssm & MSSM & \rmssm \\
\hline
$\frac{5}{3}\alpha_1$ & 0.0396 & 0.0398 & 0.0169 & 0.0169 \\
$\alpha_2$            & 0.0386 & 0.0402 & 0.0338 & 0.0338 \\
$\alpha_3$            & 0.0387 & 0.0396 & 0.1177 & 0.1171 \\
\hline
$Y_\tau$ & 0.0356 & 0.0371 & 0.0527 & 0.0527 \\
$Y_b$    & 0.0337 & 0.0343 & 0.0937 & 0.0937 \\
$Y_t$    & 0.6356 & 0.6388 & 0.9598 & 0.9598 \\
\hline
$A_\tau / Y_\tau$  & 500 & 500 & 168    & 182    \\
$A_b / Y_b$        & 500 & 500 & --1131 & --1170 \\
$A_t / Y_t$        & 500 & 500 & --853  & --861  \\
\hline
$\mu$    & 500 & 500 & 761   & 805  \\
$B\mu$   & 500 & 500 & --247 & --248 \\
\hline
$M_1$ & 500 GeV & 500 GeV& 210  GeV & 210 GeV \\
$M_2$ & 500 GeV & 500 GeV& 415  GeV & 414 GeV \\
$M_3$ & 500 GeV & 500 GeV& 1290 GeV & 1308 GeV \\
\hline
$m_{H_1}$      & 500 GeV& 500 GeV& 604 GeV   & 598 GeV  \\
$\pm|m_{H_2}|$ & 500 GeV& 500 GeV& --738 GeV & --782 GeV \\
\hline
$m^2_{udL}$& 500 GeV& 500 GeV& 1282 GeV& 1284 GeV\\
$m^2_{csL}$& 500 GeV& 500 GeV& 1281 GeV& 1284 GeV\\
$m^2_{tbL}$& 500 GeV& 500 GeV& 1155 GeV& 1163 GeV\\
\hline
$m^2_{uR}$& 500 GeV& 500 GeV& 1242 GeV& 1252 GeV\\
$m^2_{cR}$& 500 GeV& 500 GeV& 1242 GeV& 1252 GeV\\
$m^2_{tR}$& 500 GeV& 500 GeV& 963  GeV& 984  GeV\\
\hline		
$m^2_{dR}$& 500 GeV& 500 GeV& 1237 GeV& 1249 GeV\\
$m^2_{sR}$& 500 GeV& 500 GeV& 1237 GeV& 1249 GeV\\
$m^2_{bR}$& 500 GeV& 500 GeV& 1234 GeV& 1247 GeV\\
\hline
$m^2_{{\nu_e}L}$&    500 GeV& 500 GeV& 612 GeV& 607 GeV\\
$m^2_{{\nu_\mu}L}$&  500 GeV& 500 GeV& 612 GeV& 607 GeV\\
$m^2_{{\nu_\tau}L}$& 500 GeV& 500 GeV& 612 GeV& 606 GeV\\
\hline
$m^2_{eR}$&     500 GeV& 500 GeV& 536 GeV& 532 GeV\\
$m^2_{\mu R}$&  500 GeV& 500 GeV& 536 GeV& 532 GeV\\
$m^2_{\tau R}$& 500 GeV& 500 GeV& 534 GeV& 531 GeV\\
\end{tabular}
\end{ruledtabular}
\end{table}

\begin{figure*}
\includegraphics{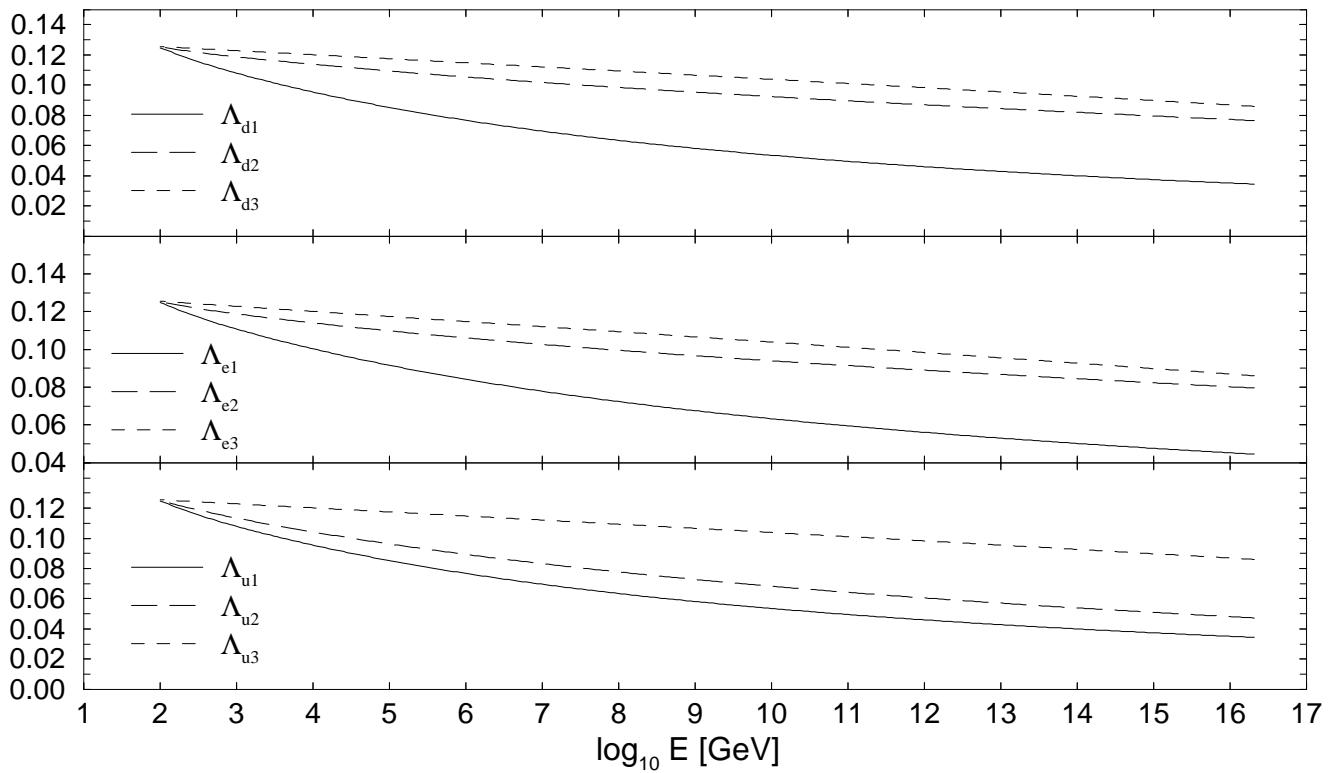}
\caption{\label{fig1} $R$-parity violating Yukawa couplings
evolution. All were set to the same value at the $m_Z$ scale and
evaluated using the 2--loop \rmssm \ RGE to the $m_{GUT}$ scale.}
\end{figure*}


The relevant 1-- and 2--loop equations for MSSM can be found in
\cite{MartinVaughn} with \rmssm \ corrections in \cite{ADedesD}. Our
procedure of runnig the parameters consists of few steps. We take into
account mass thresholds where SUSY particles start to contribute
\cite{kanekolda} and use the SM 2--loop renormalization group equations (RGE)
below appropriate threshold and 2--loop \rmssm \ RGE above it. Initially all
the thresholds are set to 1 TeV and are dynamically modified during the
running of mass parameters. After evolving the dimensionless couplings from
the electroweak scale $m_Z$ up to $m_{GUT} \sim 10^{16}$ GeV, we unify the
masses of gauginos, sfermions and squarks to be $m_0 = 500$ GeV. The
trilinear soft couplings at $m_{GUT}$ are set according to formula
$$
\mathbf{A}_i = A_0 \mathbf{Y}_i,
$$
with $A_0=500$. Next, all the quantities are evolved down to $m_Z$. The
numbers used in simulation serve as an example. Our investigation
showed, that the same conclusions can be drawn for a wide range of $m_0$ 
and $A_0$.

The values of Yukawa couplings {\bf Y} at $m_Z$ are given by lepton and
quark mass matrices $M$ \cite{ChoMisiak}
\begin{eqnarray}
M_U &=& \langle H_2^0 \rangle \mathbf{S}_{U_R} \mathbf{Y}_U^T
\mathbf{S}_{U_L}^\dagger, \nonumber \\
M_D &=& \langle H_1^0 \rangle \mathbf{S}_{D_R} \mathbf{Y}_D^T
\mathbf{S}_{D_L}^\dagger, \\
M_E &=& \langle H_1^0 \rangle \mathbf{S}_{E_R} \mathbf{Y}_E^T
\mathbf{S}_{E_L}^\dagger, \nonumber
\end{eqnarray}
where $\langle H_1^0 \rangle$ and $\langle H_2^0 \rangle$ are the
neutral Higgs vacuum expectation values. {\bf S} matrices perform
diagonalization so that one obtains eigenstates in the mass
representation. However, we had to choose some low energy values for
the remaining $R$-parity violating couplings $\mathbf \Lambda$. There is
no hint concerning their actual numerical values except that they should
be small, since lepton number violating processes are very rare. We
checked several possibilities and found, that in order to obtain
physically reliable results, all these couplings should obey
\begin{equation}
(\mathbf{\Lambda})_{ii} \le 0.015, \qquad (i=1,2,3),
\end{equation}
otherwise high vacuum expectation values are generated from the down
quark mass evolution. What is more, values of other parameters nearly
did not change for various $\mathbf{\Lambda}$'s. From \cite{WodSimkovic}
we can make use of the following relation
\begin{equation}
(\mathbf{\Lambda}_{D^1})_{11} \le \xi
\left(\frac{10^{24}y}{T_{1/2}^{0 \nu 2 \beta -exp}}\right)^{-1/4}.
\end{equation}
Using the $\xi$ parameter for $^{76}$Ge isotope for the unification
mass 500 GeV, and inserting the experimentally known half-life given by the
Heidelberg--Moscow collaboration \cite{0nu2beta} $T_{1/2}^{0 \nu 2
\beta(HMexp)} = (0.8 - 18.3) \times 10^{25}y$, we obtain
\begin{equation}
(\mathbf{\Lambda}_{D^1})_{11} \le 0.275,
\end{equation}
which is much weaker than our constraintmit (8). If one uses limit (8) the
half-life to be of the order of $\sim 1.3 \times 10^{30}y$ which is rahter
an overestimation. On the other hand, the newest data
from the IGEX Collaboration \cite{igex} sets the lower bound on the
half-life in germanium to be $T_{1/2}^{0 \nu 2 \beta(IGEXexp)} > 1.57 \times
10^{25}y$ without bounds from above.

Values for various parameters in both 1--loop MSSM and 2--loop \rmssm \
cases are given in Tab.\ref{tab1}. The $SU(3)_c \times SU(2)_L \times
U(1)_Y$ dimensionless coupling constants $g_i$ are represented here in
the usual way by $\alpha_i = g_i^2 / 4\pi$. Whenever the parameter
has a matrix form, it has been diagonalized and the entry \{33\} is
listed in the table. Masses are given respectively for gauginos,
left-handed Higgs doublets, left-handed squark doublets, right-handed
up- and down-type squark singlets, left-handed slepton doublets, and
right-handed slepton singlets, with respect to the $SU(2)$ group. 

In general, as expected, all parameters are changed by at most few percent.
One should keep in mind that the new results include two corrections: the
2--loop terms and the terms connected with $R$-parity violation. Our
investigation shows, that the differences are mainly driven by the 2--loop
corrections, whereas the $R$-parity terms change the results on the third
decimal place. 

This situation can be understood from the evolution of $R_p$-violating
couplings. In Fig.\ref{fig1} an example of running of $\Lambda$ coupling
constants is shown. It is essential, that their values decrease with
increasing energy. As a consequence the contributions coming from $\Lambda$'s
can be safely neglected in actual calculations. It is, however, a surprising
result, since normally one would expect to see exotic processes at high
energies. Fig.\ref{fig1} clearly shows, that to observe a lepton number
violating process, like the neutrinoless double beta decay, one should search
for it in low temperatures rather than in accelerators.

\section{Final remarks}

In conclusion, we have presented the spectrum of supersymmetric particles and
coupling constants using the 1--loop MSSM and 2--loop \rmssm \
renormalization group equations. Inclusion of the $R$-parity violating terms
does not change the results significantly. On the other hand we confirmed the
well known rule that the 2--loop corrections should be used in all cases,
where quantitative rather than qualitative features are important.

\begin{acknowledgments} This research was partially supported through a
European Community Marie Curie Fellowship. One of us (MG) would like to thank
the staff of Marie Curie Trainig Site, located in the European Center for
Theoretical Studies in Nuclear Physics and Related Areas (ECT*) in Trento,
Italy, for kind hospitality during Summer 2002.

\underbar{Disclaimer}: The authors are solely responsible for the information
communicated, published or disseminated. It does not represent the opinion of
the Community. The Community is not responsible for any use that might be
made of data appearing therein. \end{acknowledgments}

\bibliography{art1}
\end{document}